\documentclass[12pt]{article}
\usepackage{amsmath}
\usepackage[dvips]{graphicx}
\usepackage{amsfonts}
\usepackage{amssymb}
\usepackage{amsmath}
\usepackage{axodraw}
\def\k{\mathbf{k}}
\def\p{\mathbf{p}}
\def\q{\mathbf{q}}
\def\b{\mathbf{b}}
\def\a{\mathbf{a}}

\def\bet{\boldsymbol{\beta}}
\def\d{\partial}

\def\Pm{\hbox to 5pt {$I\!\!\!\!P$}}
\def\Rd{\hbox to 5pt {$I\!\!\!\!R$}}

\begin{document}
\title{\bf
Anisotropic flows from initial state \\ of a fast nucleus} 
\author{K.G. Boreskov, A.B. Kaidalov, O.V. Kancheli \\[3mm]
{\it Institute for Theoretical and Experimental Physics,} \\
{\it Moscow, 117218 Russia}}

\date{}
\maketitle
\begin{abstract}
We analyze azimuthal anisotropy in heavy ion collisions related to
the reaction plane in terms of standard reggeon approach and find
that it is nonzero even when the final state interaction is
switched off. This effect can be interpreted in terms of partonic
structure of colliding nuclei. We use Feynman diagram analysis to
describe details of this mechanism. Main qualitative features of
the appropriate azimuthal correlations are discussed.
\end{abstract}

\section*{Introduction}
\label{Intro}

Investigations of azimuthal anisotropy of secondary particle
distributions in heavy ion collisions is used as an effective tool
for analysis of dynamical mechanisms of these processes.
Anisotropic flows and, in particular, the elliptic flow
$v_2\equiv\langle\cos2\phi\rangle$ are considered usually as an
important source of information about  properties of  a dense
state of hadronic matter formed in the anisotropic overlap region
of colliding nuclei \footnote{ See e.g. \cite{Olli92} -
\cite{starflow}.
}%
. Crucial point is the relation of the anisotropy to the reaction
plane orientation (azimuthal correlations not related to this
orientation are called non-flow correlations). The dependencies of
flow coefficients on c.o.m. energy, centrality, rapidity and
transverse momentum of secondaries are ascribed usually to
anisotropic properties of strongly interacting matter at early
times after collision. It is assumed that anisotropic flows are
due to final state interactions and in hydrodynamic models
originate from pressure gradients connected to original asymmetry
in the configuration space for noncentral collisions. Practically
all existing models for anisotropic flows use classical
description of processes.

In this paper we use quantum relativistic approach, based on
analysis of Feynman diagrams and reggeon diagrams for a study of
anisotropic flows in collisions of hadrons and nuclei at high
energies. We show that a part of azimuthal anisotropy can be
related to properties of \emph{initial} state of a fast nucleus.
Namely, we analyze azimuthal correlations present in partonic
configurations of fast colliding nuclei. Thus a part of the
observed azimuthal asymmetry of final particles can come from
parton correlations in initial state. We demonstrate how
anisotropic flows in this case are related to the anisotropy of
the overlap region.

It is worth to point out that, though it is not possible to
measure in general both transverse momenta and transverse
coordinates of partons, but for fast moving heavy nucleus impact
parameters of nucleons and fast partons can be fixed with a good
accuracy. Therefore for colliding nuclei the radial directions in
every point of transverse plane with respect to c.o.m. are well
defined. Distributions of partons at each point are anisotropic in
general. The collision process (if it is not absolutely central)
selects partonic configurations  asymmetric in impact parameter
plane. This transforms the initial `radial' asymmetry in colliding
nuclei to the azimuthal asymmetry of produced particles momenta
with respect to the collision plane.

The simplest way to study such phenomena theoretically is to use
the reggeon diagrams for inclusive cross sections, where all
details of parton interactions are encoded  in vertices  and Regge
trajectories -- and we follow this way in the paper.

The paper \footnote{~This paper is an extended version of a talk
given at Session of Nuclear Physics Division of Russian Academy
of Sciences in November 2007.
}
is organized as follows. In Sec.\,\ref{momentum} we calculate a
contribution to $v_2$ due to standard multiperipheral production
described by reggeon diagram for inclusive production. The
calculation is carried out in the momentum representation and
gives $v_2$ as a function of impact parameter $\b$ and particle
transverse momentum $\p_t$. A simple phenomenological
parameterization is used for the vertex of a detected particle
emission.
Then in Sec.\,\ref{impact} we give a qualitative interpretation of
the results in terms of the impact plane analysis. It is shown
that the obtained azimuthal asymmetry can be ascribed to specific
correlations presented at the level of partonic configuration of
wave function of the fast nucleus. It exists without any final
state interactions, which are usually considered as an obligatory
condition for appearance of correlations related to the reaction
plane orientation. We explain how these primary correlations
manifest themselves as the elliptic flow due to anisotropy of the
overlap region.
In Sec.\,\ref{vertex} we make an estimate of the anisotropy
structure of the inclusive vertex in simple multiperipheral models
and show that they naturally result in azimuthal dependence.
Possible mechanisms enhancing the effect are discussed.
In Sect.\,\ref{directed} we discuss generation of the directed
flow effect in the same model. This effect is produced by
nondiagonal in reggeons $RP$ vertex and therefore it is
concentrated at edges of rapidity space.
In Sec.\,\ref{two-particle} we discuss contribution to 2-particle
azimuthal correlations in reggeon approach and suggest possible
tests of this mechanism.
Sec.\,\ref{rescatt} contains discussion of multipomeron
interactions which can have crucial influence on magnitude of the
effect and on its atomic number and impact parameter dependencies.
In Sec.\,\ref{numerical} numerical estimates are presented to
illustrate main qualitative features of the model.
In Sec.\,\ref{conclusion}  we briefly summarize our main results.

\section{Momentum representation}
\label{momentum}

\begin{figure}[h]
\begin{center}
\includegraphics[height=40mm]{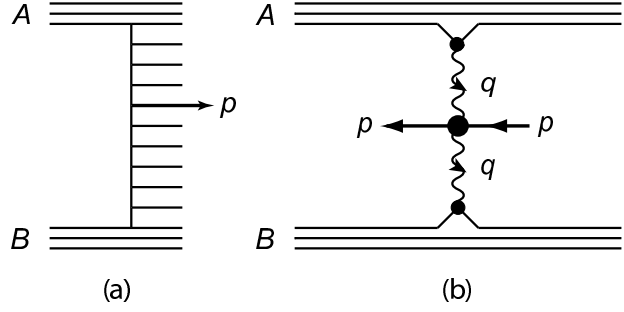}
\caption[in-reggeon]{Reggeon diagram for one-particle inclusive production}
\label{incl1}
\end{center}
\end{figure}

Let us explain first the mechanism under consideration using reggeon diagrams.
A simplest way to produce secondary particles at high energy is
given by multiperipheral diagram of Fig.\ref{incl1}a. Inclusive
cross section is described by the reggeon diagram of
Fig.\ref{incl1}b which corresponds to squared multiparticle
amplitude with fixed momentum $p$ of a single detected particle.

Single particle inclusive cross section is related to the amplitude $G(\q,\p)$
of the diagram of Fig.\ref{incl1}b at zero transfer momentum $\q=0$:
\begin{align}
\label{incl}
F_{incl}(\p_t;y)\equiv\frac{d\sigma}{d^2\p_t dy} \sim G(\q=0;\p_t,y) ~,
\end{align}
where $p_t$ and $y$ are transverse momentum and rapidity of the registered particle.
We, however, need to calculate a contribution of this diagram at various
values of $\q$ in order to find a correlation between transverse momentum $\p_t$ and
impact parameter value for colliding nuclei, $\b$. This relation is given by Fourier
transform of $G(\q;\p_t,y)$
\begin{align}
\label{incl_b}
        \widetilde{F}_{incl}(\p_t,y;\b)\sim
        \widetilde{G}(\b;\p_t,y)=
        \int\ \frac{d^2\q}{(2\pi)^2}\ e^{i\b\q}\ G(\q;\p_t,y) ~.
\end{align}
Here
\begin{align}
\label{incl_q}
        G(\q;\p_t,y)=F_{A}(q)D(\q;y)\gamma(\q,\p_t)D(\q;Y-y) F_{B}(q) ~,
\end{align}
where $F_{A,B}$ are 2-dimensional form factors of colliding nuclei which are
Fourier transforms of their 2-dimensional profiles, $T_{A,B}(b)$,
\begin{align}
    F_{A,B}(q) = \int d^2b\ e^{-i\q\b}T_{A,B}(b) ~,
    \qquad F_{A}(0) = A, \quad F_{B}(0) = B,
\end{align}
and functions $D(\q;y),D(\q;Y-y)$ are connected
to s-channel discontinuities of reggeon amplitudes (``cut reggeons'')
\footnote{We use an exponential parameterization of reggeon vertex
$g_N \exp(-R_0^2 \q^2)$ and linear Regge trajectory
$\alpha(\q^2)=\alpha(0)-\alpha\,^\prime \,\q^2$.}
\begin{align}
    D(\q;y)\equiv 2\,\mathrm{Im} f(\q;y)= g_N e^{-(R_0^2 +\alpha\,^\prime y)q^2}
    e^{[\alpha(0)-1]y}~.
\end{align}
An inclusive vertex of the reggeon diagram $\gamma(\q,\p_t)$ depends on 2-dimensional
momenta $\p_t$ and $\q$.
For estimates we will use the phenomenological parametrization
\begin{align}
    \label{gamma_exp}
    \gamma(\q,\p_t) = \gamma_0 \exp\{-r_q^2 \q^2
    - r_p^2 \p_t^2-\epsilon r_0^4 (\p_t\q)^2\} ~.
\end{align}
Here dimensional coefficients $r_q,r_p$ are expected to be of
typical hadronic sizes about $r_0 = 1~\mathrm{fm}$, and $\epsilon$
of order of 1.

The dependence on variable $(\p_t\q)$ is crucial
for appearance of azimuthal correlations typical for elliptic
flow. Note that the terms proportional to the odd powers of
$\p_t\q$ should be absent if both reggeons have positive
$C$-parity (like pomerons) because the vertex should be even under
the change $\p_t\rightarrow -\p_t$. More detailed analysis of
structure and value of the vertex $\gamma(\q,\p_t)$ will be
performed in sections \ref{vertex},\ref{directed}.

The elliptic flow $v_2$ is defined as a second Fourier coefficient in an azimuthal
dependence of inclusive spectrum%
\begin{align}
    \frac{d\sigma}{dp_t^2 d\phi} \propto
    1 + \sum_1^\infty 2v_n(b;p_t,y)\cos(n\phi)    ~,
\end{align}
where $\phi$ is an azimuthal angle between impact parameter $\b$ which determines
the reaction plane and transverse momentum $\p_t$.
Thus,
\begin{align}
\label{v2}
    v_2 (b;p_t,y) \equiv \frac{C_2}{C_0}
    = \frac{\int d\phi\ cos(2\phi)\ G(\b;\p_t,y)}{\int d\phi\ G(\b;\p_t,y)}
    ~.
\end{align}
For the vertex function parameterized by eq.\eqref{gamma_exp} one gets
\begin{align}
    C_n(b;p_t,y) &= \int qdq J_n(bq) F_{A}(q^2)F_{B}(q^2)D(q;y)D(q,Y-y)
    c_n(q;b,p_t) ~, \quad n=0,2,
\end{align}
with
\begin{align}
    C_0 (q;b,p_t) &= I_0(\epsilon r_0^4 p_t^2q^2/2)\ e^{-\epsilon r_0^4 p_t^2q^2/2} ~, \\ \nonumber
    C_2 (q;b,p_t) &= I_1(\epsilon r_0^4 p_t^2q^2/2)\ e^{-\epsilon r_0^4 p_t^2q^2/2} ~.
\end{align}
where $I_0,I_1$ are the modified Bessel functions.

 To estimate an order of the effect let us
first make the calculation with the approximate Gaussian
parameterization of nuclear form factors $F_{A,B}(q^2)$:
\begin{align}
\label{dens_gauss}
    F_{A,B}(q^2) = \exp(-R_{A,B}^2 q^2/4) ~.
\end{align}
Then
\begin{align}
\label{v2_gauss}
    v_2(b;p_t,y) = \frac{I_1(\alpha/2)}{I_0(\alpha/2)}
\end{align}
with
\begin{align}
    \alpha &= \frac{\epsilon r_0^4 p_t^2 b^2}{R^2(R^2+4 \epsilon r_0^4 p_t^2)}
    ~,\qquad     R^2 = R_A^2 +R_B^2 + R_{regge}^2(Y)~.
\end{align}
For ~$\alpha \ll 1$ we have ~:
\begin{align}
\label{v2approx}
v_2 \simeq \frac{\alpha}{4}\simeq \epsilon\,(r_0 p_t)^2\,\left(\frac{r_0}{R}\right)^2\,
        \left(\frac{b}{2 R}\right)^2 ~.
\end{align}
Thus, $v_2$ is proportional to the ``ellipticity'' $\epsilon$ of the
inclusive vertex $\gamma$. The sign of $v_2$ is the same as the one of
$\epsilon$ which can be both positive and negative depending on
dynamical structure of the inclusive vertex (see discussion in
Sect.\ref{vertex}). The formula \eqref{v2approx} reveals an
increase of $v_2$ with transverse momentum $\p_t$ and impact
parameter $b$.
Results of numerical calculations are presented in sec.\ref{numerical}.

\section{Impact parameter representation}
\label{impact}

\begin{figure}[h]
\begin{center}
\includegraphics[height=60mm]{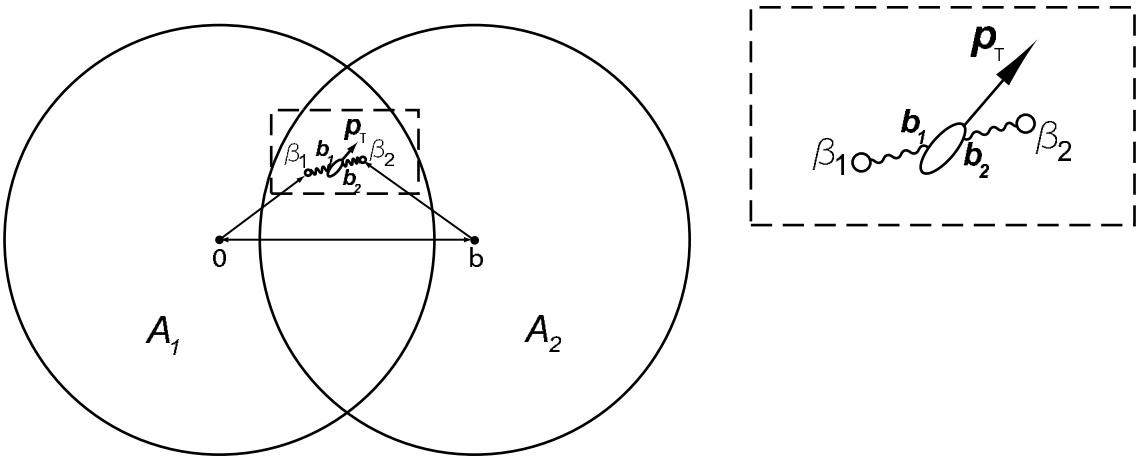}
\caption[in-reggeon]{Interaction of two nucleons in impact plane}
\label{impact2}
\end{center}
\end{figure}

Presentation of formulas in terms of impact parameters allows us
to give more instructive interpretation of the results.
Eqs.\eqref{incl_b},\eqref{incl_q} in these variables have the
form:
\begin{align}
\label{incl2}
    G(\b;\p_t,y)
    \sim&
     \int\! d^2\bet_1 \!\int\! d^2\b_1
     \!\int\! d^2\b_2 \!\int\! d^2\bet_2\ T_{A}(\bet_1)\ t_1(\b_1 - \bet_1;y)\
     \nonumber\\
    & \times \Gamma(\b_2-\b_1,\p_t)\ t_2(\bet_2 - \b_2;Y-y)\ T_{B}(\b-\bet_2) ~.
\end{align}
Here (see Fig.\ref{impact2}) $\bet_1$ and $\b-\bet_2$ are the
positions of participant nucleons of the nuclei $A$ and $B$ with
nuclear profiles $T_{A}$ and $T_{B}$, $t(\bet;y)$ is a Fourier
transform of reggeon propagator $D(\q;y)$ and $\Gamma(\b;\p_t)$ is
a Fourier transform of the inclusive vertex $\gamma(\q,\p_t)$:
\begin{align}
     \Gamma(\b;\p_t) =
        \int\! \frac{d^2\q}{(2\pi)^2}\ e^{i\q\b} \gamma(\q,\p_t) ~.
\end{align}
For the exponential model approximation \eqref{gamma_exp}, one has
\begin{align}
\label{gamma_exp_b}
        \Gamma(\b;\p_t) = \frac{\pi e^{-r_p^2 p_t^2}}{\sqrt{r_q^2 (r_{q}^2
        +\epsilon r_0^4 p_t^2)}}
        \exp\left(-\frac{b_x^2}{4(r_{q}^2+\epsilon r_0^4 p_t^2)}
        -\frac{b_y^2}{4r_q^2}\right)~,
\end{align}
where $b_x$ is a component of $\b$ along $\p_t$ and $b_y$ in the transverse direction.
In order to reveal a structure of Eq.\eqref{incl2} it can be written symbolically as
a number of convolutions with notation $f\otimes g$ for convolution
of functions $f$ and $g$:
\begin{align}
\label{conv0}
    F_{incl}(\p_t,y;\b) \sim T_A\otimes t(y)\otimes \Gamma\otimes t(Y-y)\otimes T_B ~.
\end{align}
Convolutions in eq.\eqref{incl2} can be done in arbitrary order, so it is equivalent to
\begin{align}
\label{conv1}
    F_{incl}(\p_t,y;\b) \sim T_{int}\otimes \Gamma  =
    \int d^2 \a\, T_{int}(\b-\a)\, \Gamma (\a;p_t) ~.
\end{align}
Here the function $T_{int}$ which characterize a region of interaction of nuclei in
$\b$ space is a convolution of the nuclear overlap
\footnote{
~To avoid confusion let us stress that $T_{AB}$ describes how the number of
nucleons from colliding nuclei which overlap in the impact plane depends on collision
impact parameter $b$. It is an isotropic function not depending on the vector
$\b$ direction.
}
$T_{AB}=T_{A}\otimes T_{B}$ and combined reggeon amplitude $t_{regge}$:
\begin{align*}
T_{int} = T_{AB} \otimes t_{regge} ~,
\end{align*}
where
\begin{align*}
t_{regge}(b;Y) &=t(y)\otimes t(Y-y)
= \frac{g_N^2}{4\pi R_{regge}^2(Y)} \exp{\left(-\dfrac{b^2}{4R_{regge}^2(Y)}\right)}~,\\
R_{regge}^2(Y) &= 2 R_0^2 + \alpha\,^{\prime} Y ~,
\end{align*}
describes the regge-interaction amplitude of nucleons from colliding nuclei.

For Gaussian parameterization of nuclear profiles $T_{A}$, $T_{B}$
all convolutions have the Gaussian form again, resulting in
the same final answer \eqref{v2_gauss}.

In the impact parameter representation it is easy to see an origin of the correlation between
directions of $\b$ and transverse momentum $\p_t$.
Registration of the inclusive particle is realized by nonlocal probe which structure
is determined by the inclusive vertex $\Gamma(\b;\p_t)$.
This vertex has an elliptic anisotropy in $b$-plane along the direction
of $\p_t$ (see examples in Sec.\ref{vertex}).
Due to this anisotropy the convolution of the overlap function $T_{int}$ and $\Gamma$
is sensitive to gradients of nuclear densities.

The size of the probe range is small in comparison to nuclear size
$R_{A}\sim{R_B}\sim A^{1/3}r_0$, so it is possible
to calculate the integral \eqref{conv1} expanding the smooth distribution $T_{int}(\b-\a)$
over small corrections $a_x/R_A,a_y/R_A$.
After integration over $a_x, a_y$ we get explicitly anisotropic expression for the
inclusive cross section (recall that the axis $x$ is chosen along the momentum $\p_t$).
Omitting higher order terms in $a_i/R_A$ we have in the approximation
\eqref{gamma_exp},\eqref{gamma_exp_b}
\begin{align}
\label{incl_imp}
    F_{incl}\sim T_{int}(b^2)+2(2r_{q}^2+\epsilon\, r_0^4 p_t^2)\frac{dT_{int}(b^2)}{d(b^2)}
    +4 \frac{d^2T_{int}(b^2)}{d(b^2)^2}\left[(r_{q}^2
    +\epsilon\, r_0^4 p_t^2)b_x^2+r_{q}b_y^2\right] ~,
\end{align}
where derivatives are taken over the variable $b^2$.
Thus, for the elliptic flow coefficient we have
\begin{align}
\label{v2T}
    v_2 \simeq \epsilon\, \frac{r_0^4 T_{int}^{\,\prime\prime}(b^2)}{T_{int}(\b^2)}\,
    \p_t^2\,\b^2 ~.
\end{align}
The second derivative of $T_{int}(\b^2)$ contains a small parameter of order of
$(r_0/R_A)^4$ which appears explicitly in Eq.\eqref{v2approx}.

This formula can be interpreted in terms of parton probe. The local probe
feels only parton density.
An information about density gradients and its higher derivatives
can be obtained by means of nonlocal probes with typically hadronic range.
Being restricted only with lower multipoles a structure of the probe can be written as
\begin{align}
\label{dipole}
    \Gamma(\a;\p) \sim \left(1 + \bar{\delta} p_i \d_i +
    (\beta \delta_{ij} +\bar{\epsilon} p_i p_j)\d_i\d_j + \dots\right) \delta^{(2)}(\a) ~,
\end{align}
where $\d_i=\d/\d a_i$ and
$\bar{\delta}\equiv \delta r_0^2$, $\bar{\epsilon}\equiv\epsilon r_0^4$
are dimensional coefficients of typical hadronic range.
The first term measures parton density as well as the isotropic one with $\delta_{ij}$,
the second term feels gradients, and the $\epsilon$-term gives an information about
ellipticity of partonic distribution at the point of production of registered particle.
So the distribution over transverse momenta turns out to be correlated
with impact parameter $\b$:
\begin{align}
\label{conv2}
    F_{incl}(\p_t;\b) &\sim T_{int}\otimes \Gamma  =
    \int d^2 \a T_{int}(\b-\a) \Gamma (\a;\p_t)   \nonumber \\
    &\approx T_{int}(b^2) + \bar{\delta} p_{t\,i} \d_i T_{int}(b^2)
    + \bar{\epsilon} p_{\,t\,i} p_{\,t\,j}\d_i\d_j T_{int}(b^2) \nonumber\\
    &\approx  T_{int}(b^2) +  2\bar{\delta} (\p_{\,t} \b) T_{int}^{\,'}(b^2) +
    4\bar{\epsilon} (\p_t \b)^2 T_{int}^{\,''}(b^2) ~,
\end{align}
where $T^{'}$ denotes the derivative over $b^2$. This reproduces the elliptic flow
coefficient $v_2=\epsilon (\p_t \b)^2 T_{int}^{''}(b^2)$ (cf. with \eqref{v2T}).
The $\delta$-term corresponds to the first flow coefficient
$v_1=\delta (\p_t\b)\, T_{int}^{'}(b^2)$ (directed flow), which we discuss
in Sec.\,\ref{directed}.

Let us stress that partonic interpretation of reggeon diagrams depends
on chosen Lorentz frame. Reggeon exchanges are highly non-local in space-time
due to their complicated multiparticle internal structure.
From this viewpoint the fast hadron
or nucleus is essentially a multiparton state of multiperipheral configuration
(Fig.\ref{incl1}a represents the simplest example). The particles of such configuration
are ordered in their rapidities and only most slow particles interact with a target.
This means that in the lab frame related to the nucleus $B$ the nucleus $A$
is a multiparton state containing the detected particle with rapidity $y$ and
transverse momentum $\p_t$ among others. This state has already an anisotropy in
partonic distributions due to correlation between parton transverse momentum $\p_t$
and its position $\b_1$ in impact parameter plane. If nuclear collision is central
then the overlap region is isotropic and after integration over $b_1$ anisotropy
disappears. At non-central collision, however, the overlap region has anisotropic
almond shape and integration over this region keeps anisotropic distribution in $\p_t$.

\section{Inclusive vertex}
\label{vertex}

\begin{figure}[h]
\begin{center}
\includegraphics[height=40mm]{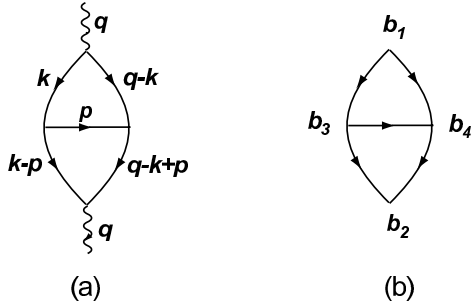}
\caption[in-reggeon]{Simplest inclusive reggeon vertex}
\label{vert1}
\end{center}
\end{figure}

Let us estimate in simple models an anisotropy of inclusive $\p_t$ probe related to the
vertex $\gamma(\q,\p_t)$. If multiperipheral ladder of Fig.\ref{incl1} consists of
scalar particles with triple coupling then a structure of inclusive vertex is determined
in lowest order by Fig.\ref{vert1} where momentum $\p_t$ is fixed.
Then the vertex can be written as 2-dimensional integral
\footnote{~
There is an integration over 3-dimensional momentum $\k$ in the loop but for estimate
we shall assume that longitudinal integration is already carried out.
It introduces $t_{min}$ terms into propagators which can be replaced by change of effective
masses.
So, in what follows we use propagators as functions of 2-dimensional momenta.}
\begin{align}
    \label{gammaloop}
    \gamma(\q,\p_t) = g^2 \int d^2\k\ G(\k)G(\k - \p_t)G(\q - \k)G(\q - \k + \p_t) ,
\end{align}
where particle propagators $G$ depend on 2-dimensional momenta only.

To estimate an anisotropy effect let us consider the vertex $\gamma$ at small
$q\sim p_t \ll\mu$ (here $\mu\sim r_0^{-1}$ is a typical mass parameter):
\begin{align}
\label{gamma_expand}
    \gamma(p_t,q,z) \simeq \gamma(p_t^2) (1 - \epsilon \frac{q^2 p_t^2}{\mu^4}z^2 + \dots)  ~,
\end{align}
where $z=\cos{\phi}$ is a cosine of the angle between $\p_t$ and $\q$.
The coefficient which determines a degree of ellipticity is denoted as $\epsilon$
as we did before.
For standard perturbative behavior of propagators:
    $G(\k ) = ( \k^2 + \mu^2 )^{-1}$~~
numerical estimate gives a value $\epsilon \approx - 0.2$ for scalar couplings $g$
in the diagram \ref{vert1}. Note that such a sign corresponds to negative $v_2$
(in contrast to experimental observation).
This fact can be understood in a simple way by writing Eq.\eqref{gammaloop}
in coordinate representation (see Fig.\ref{vert1}b):
\begin{align}
    \label{gammaloopx}
    \widetilde{\gamma}(\b_{12},\b_{34}) =& g^2 \int \prod_i d^2\b_i\,
        \widetilde{G}(b_{13})\widetilde{G}(b_{32})\widetilde{G}(b_{24})\widetilde{G}(b_{41})
     \\  \nonumber
    &\qquad\times\delta^2(\b_{12}-(\b_2-\b_1))\delta^2(\b_{34}-(\b_4-\b_3))   ~,
\end{align}
where $\b_{ij}~=~\b_i-\b_j$, and the vector $\b_{12}$ is conjugated to $\q$, while
$\b_{34}$ -- to the momentum $\p_t$.
The 2-dimensional Green functions $\widetilde{G}(\b)\sim \exp{(-\mu|\b|)}$ and
therefore the integrand in \eqref{gammaloopx} depends only on sum of all lengths
$|\b_{13}|+|\b_{32}|+|\b_{24}|+|\b_{41}|$. It can be seen purely geometrically that
maximum of $\prod \widetilde{G}$ at fixed $|\b_{34}|,|\b_{12}|$ is reached in the
configuration with parallel $\b_{12}$ and $\b_{34}$. The maximal anisotropy (\,$\sim 0.3$\,)
corresponds to $|\b_{12}|/|\b_{34}|\sim 1$ and decrease strongly with change of this ratio.
Since $\gamma(\b_{12},\b_{34})$ is maximal for parallel $\b_{12}$ and $\b_{34}$, then
$\gamma(\q, \p_t)$ is maximal at parallel $\q$ and $\p_t$, i.e. $\epsilon$ is negative.
Similar arguments can be generalized to more complicated diagrams with spinless
particles -- configurations with parallel $\b_{12}$ and $\b_{34}$ are more probable.

\begin{figure}[h]
\begin{center}
\includegraphics[height=40mm]{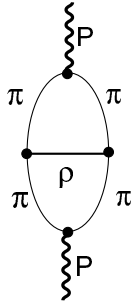}
\caption{$PP$-diagram for inclusive $\rho$-meson production}
\label{PP1}
\end{center}
\end{figure}

The value and the sign of the effect depend on dynamical structure of the inclusive
vertex $\gamma(\q, \p_t)$ -- on form of propagators
\footnote{~Note that in case of Gaussian form of propagators, $G(k) = \exp(-k^2/\mu^2)$,
the product of propagators in coordinate space does not depend on the angle resulting
in zero anisotropy, $\epsilon=0$.}
and, more important, on structure of vertex of particle emission.
Vertex properties are very sensitive to particle spin
because it gives an extra azimuthal dependence. Calculations show that if the inclusive
particle has spin 1 (see, Fig.\ref{PP1}), then the sign of $\epsilon$ is different
(in accordance with data) and its value increases.
Note that such situation is not far from reality because $\rho$-mesons give
a considerable part of multiperipheral particles and pions due to their decays
have mainly close transverse momenta.
Numerical estimate gives $\epsilon\sim 0.5$.

Properties of the inclusive vertex $\gamma(\q, \p_t)$ can be studied experimentally
in $NN$ collisions, for instance by analysis of azimuthal two-particle correlations
for hadrons in different windows of rapidity values.

Let us discuss behaviour of the inclusive vertex $\gamma(\q, \p_t)$ at large $p_t$.
In region $\p_t^2 \gg \mu^2$ it should decrease according to perturbation theory
in power-like way, $\sim (\mu/p_t)^m$, where a degree $m$ depends on dynamics.
In simple scalar theory (\ref{gammaloop}) $\gamma(\q, \p_t) \sim (\mu/p_t)^4$,
and $z$-depending coefficients decrease even faster, leading to a small ellipticity:
\begin{align*}
    |\epsilon (p_t)| =
\left|\frac{\gamma(q^2, p_t^2,z=1)- \gamma(q^2, p_t^2,z=0)}{\gamma(q^2, p_t^2,z=1)}\right|
  \sim (\mu/p_t)^2 ~ \quad \text{at~} p_t^2 \gg \mu^2 .
\end{align*}
It means that a size of the $p_t$-probe becomes much smaller than $\mu^{-1}$ and
its form becomes more isotropic
\footnote{
~ It is clearly seen in the coordinate representation (\ref{gammaloopx}),
since $b_{12}^2 \sim \mu^{-2}$, $b_{34}^2 \sim p_t^{-2} \ll \mu^{-2}$,
and the integrand depends on relative orientation of $\b_{12}$ and $\b_{34}$
only in high orders in $(\mu/p_t)$
}.

The situation is different if the vertex includes particles with spin because
in this case spin terms give an additional $z$ dependence.
As an example, in case of inclusive production of vector particles explicit
calculation gives non-vanishing vertex ellipticity,
$\epsilon (p_t) \rightarrow \rm{const}$ when $p_t^2 \gg \mu^2$.

\section{Directed flow}
\label{directed}

The mechanism applied to generation of the elliptic flow $v_2$ contributes to all
$v_n$-harmonics. As it was mentioned in Sec.\ref{momentum} there is a pecularity related
to odd harmonics -- the $PP$ reggeon diagram do not contribute there because of symmetry
of the $PP$ inclusive vertex. The odd harmonics appear for vertices which couple
different types of reggeons ($RP$ vertices where $R$ is a secondary reggeon, e.g. $\rho$).

\begin{figure}[h]
\begin{center}
\includegraphics[height=40mm]{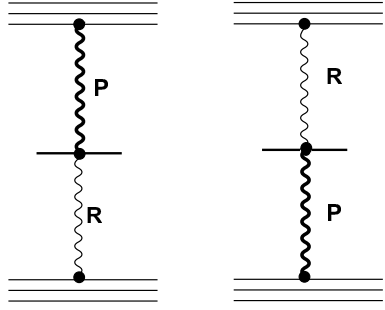}
\caption[in-reggeon]{Inclusive reggeon-pomeron diagrams contributing to $v_1$.}
\label{RP}
\end{center}
\end{figure}

Consider as an example contribution of diagrams of Fig.\ref{RP}
to the first harmonic corresponding to the directed flow $v_1$.
These diagrams contain the rapidity dependent factor which comes
from the reggeon propagator
$D_R(\q,y)\sim \exp(\Delta_\rho y)$ where $\Delta_\rho \approx 0.5$.
The diagrams $(a)$ and (b) have different sign for the term $\q\p_t$
so for the rapidity changing in the interval $[-Y,Y]$ we get a factor
$f_\rho(y)=2\exp(-\Delta_\rho Y) \sinh(\Delta_\rho y)$.

\begin{figure}[h]
\begin{center}
\includegraphics[height=40mm]{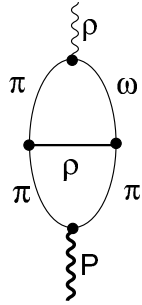}
\caption[in-reggeon]{$RP$-diagram for inclusive $\rho$-meson production}
\label{RP1}
\end{center}
\end{figure}

We have to compare the term in $\gamma^{(RP)}$ proportional to $\q\,\p_t$
with the angle independent term $\gamma(\p_t,\q=0)$
due to main diagram of Fig.\,\ref{incl1}b (see Eq.\eqref{gammaloop}).
For small $\p_t$
\begin{align}
\label{RP_vertex}
 \gamma_1^{(RP)}(\q,\p_t)/\gamma(\p_t,\q=0) \simeq
 \frac{\delta}{\mu^2} \q\, \p_t ~.
\end{align}
Calculation of the direct flow $v_1$ due to the $RP$ diagrams are similar to ones made in Sect.\ref{vertex}
and give (compare to \eqref{v2approx})
\begin{align}
\label{v1approx}
v_1  \simeq \delta \,(r_0 p_t)\,\left(\frac{r_0}{R}\right)\,\left(\frac{b}{2 R}\right)f_\rho(y) ~.
\end{align}
The numerical estimate of diagram of Fig.\,\ref{RP1} gives $\delta\sim 0.1\div 0.5$.
Qualitative behaviour of $v_1$ as a function of $y$,
$b$ and $p_t$ is in agreement with experimental data (see Sect.\ref{numerical}).

Note also that in this model the expressions for high-order flows contain at small $p_t$
the same parameters as $v_1$ and $v_2$ but in higher powers.
\begin{align}
\label{vnapprox}
v_n  \simeq \delta_n\,(r_0\, p_t)^n\left(\frac{r_0}{R}\right)^n
        \left(\frac{b}{2 R}\right)^n ~.
\end{align}

\section{Two-particle azimuthal correlations}
\label{two-particle}

We discussed hitherto correlations of transverse momentum of the
detected particle with respect to the reaction plane. The reaction plane orientation
is difficult to reconstruct from experimental data, and often
another method for studying anisotropic flow is used.
It is related with analysis of two-particle azimuthal distributions, which contain
an information on anisotropic flow. As it is seen from analysis of experimental data
the main part of these correlations is related to the $\p_t-\b$ correlations for
each particle. However, there are possible sources of
two-particle correlations unrelated to the reaction plane, so called
`non-flow' correlations.

\begin{figure}[h]
\begin{center}
\includegraphics[height=60mm]{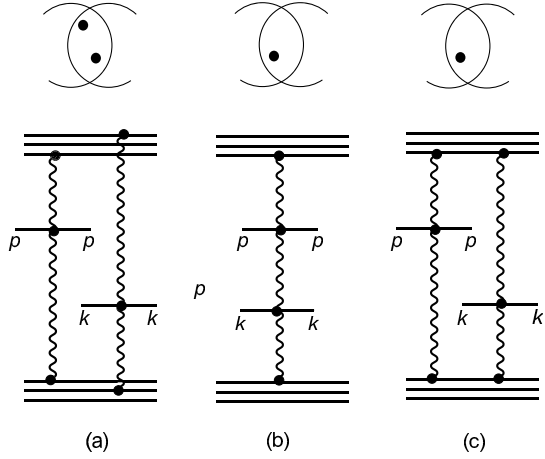}
\caption[in-reggeon]{Reggeon diagrams for two particle inclusive production.
(a) - particles p and k are emitted from different regions in transverse plane;
(b) - p and k are emitted from a single reggeon chain;
(c) - p and k are emitted from two reggeon chains attached to a single nucleon pair
in colliding nuclei.
 }
\label{f_incl2}
\end{center}
\end{figure}

Corresponding two-particle inclusive reggeon diagrams are shown in fig.\ref{f_incl2}.
To calculate anisotropic flows these diagrams should be estimated not at zero
momentum transfer $\q=0$ as for standard bi-inclusive cross section but at fixed value
of impact parameter $b$.

The contribution of the diagram \ref{f_incl2}a is factorized in the impact parameter
representation and contains factors corresponding to one-particle production
of Fig.\ref{incl1}b (see \eqref{conv2}).
\begin{align}
\label{2particles}
    F_{incl}^{(a)}(\p_t,\k_t;\b) &\sim
        \left[ T_{int}(b^2)+4\epsilon (\p_t\cdot \b)^2 T_{int}^{''}(b^2)\right]
        \left[ T_{int}(b^2)+4\epsilon (\k_t\cdot \b)^2 T_{int}^{''}(b^2)\right]~.
\end{align}
This results in factorized formula:
\begin{align}
\label{incl_two}
    \left\langle e^{2i (\phi_{p_t}-\phi_{k_t})}\right\rangle
    = \frac{\int d^2\p_t d^2\k_t\, e^{2i (\phi_{p_t}-\phi_{k_t})}\, F_{incl}(\p_t,\k_t;\b)}
    {\int d^2\p_t d^2\k_t \, F_{incl}(\p_t,\k_t;\b)}
    = v_2(p_t) v_2(k_t)  ~.
\end{align}
The contribution of diagram \ref{f_incl2}b is given by a convolution of $T_{int}$ with two probes:
\begin{align}
    F_{incl}^{(b)}(\p_t,\k_t;\b)
    &=  \int d^2 \a_1 d^2 \a_2\,
    T_{int}(\b-\a_1 - \a_2) \Gamma (\a_1;\p_t)  \Gamma (\a_2;\k_t)  ~.
\end{align}
This distribution (in the dipole approximation \eqref{dipole} with $\alpha=0$)
will contain higher derivatives of $T_{int}$ over $b^2$:
\begin{align}
    F_{incl}^{(b)}(\p_t,\k_t;\b) \sim
    T_{int} &+ 8\epsilon^2 (\k_t\cdot \p_t)^2 T_{int}^{''}
    + 32\epsilon^2 (\k_t\cdot \p_t)(\p_t\cdot \b)(\k_t\cdot \b)T_{int}^{(3)} \\ \nonumber
    &+ 8\epsilon^2 (\p_t\cdot \b)^2 (\k_t\cdot \b)^2 T_{int}^{(4)}
    ~.
\end{align}
All $\epsilon^2$-terms are at $b\sim R$ of the same order of value, $O(1)$,
and are small compared to the main term in Eq.\eqref{2particles},
which is of order $O(R^2)\sim A^{4/3}$.
So extra unfactorized correction to Eq.\eqref{incl_two} due to diagram \ref{f_incl2}b
is inessential.

The situation is different for diagram 4c. Its isotropic part has a smallness of
order $A^{-4/3}$ in comparison with isotropic part of $F_{incl}^{(a)}$.
The anisotropic part contains a `non-flow' term $(\p_t\cdot\k_t)^2$ which contains
the overlap function $T_{AB}$ (not its derivatives). As a result its contribution
to \eqref{incl_two} is of the same order as due to factorized structure
$(\p_t \b)^2(\k_t \b)^2$ from Eq.\eqref{2particles}. This contribution violates
factorization in Eq.\eqref{incl_two} for two-particle correlation.
Note that the `flow' and `non-flow' structures have different dependence on impact
parameter.

\section{Role of multipomeron interactions}
\label{rescatt}

In this section we try to estimate contributions to azimuthal asymmetry coming
from multipomeron interactions.
The simplest reggeon diagram of Fig.\ref{incl1}b gives main contribution to
inclusive cross section. Contributions of multipomeron diagrams of Glauber type
cancel each other because of the AGK rules \cite{AGK}
\footnote{~ The cancellation occurs between diagrams with different numbers
of reggeon chains (cut reggeons) considered together with sign-alternating
absorptive corrections (uncut reggeons). Note that consistent account for
absorptive corrections
is an important feature of the reggeon approach which differs it from all
classical considerations of multiparticle production.}.
Note that in `event-by-event' analysis the number of secondary particles is fixed
for every individual event, and the AGK rules may not be fulfilled. However,
all extra interactions change both isotropic and anisotropic parts of the amplitude
to the same extent, and therefore do not change the flow coefficients $v_n$.

In reggeon theory with $\alpha_P(0)>1$ the number of pomeron exchanges increases
with energy, so a role of inter-pomeron interactions grows. Detailed analysis
of multipomeron effects will be published elsewhere, and here we
will discuss briefly physical reasons for possible increase of azimuthal anisotropy.

When energy increases a longitudinal length of the nucleon tube with fixed
transverse position is contracted and multiperipheral fluctuations connected to
each of fast nucleons overlap each other in longitudinal size. Therefore
a probability for particles (partons) of different chains (related to different
nucleons) to interact increases. Two chains can either fuse to a single chain
(triple pomeron interaction) or rescatter without changing a whole chain number
($2$-$2$ or, more generally, $n$-$n$ pomeron vertices).
\begin{figure}[h]
\begin{center}
\includegraphics[height=60mm]{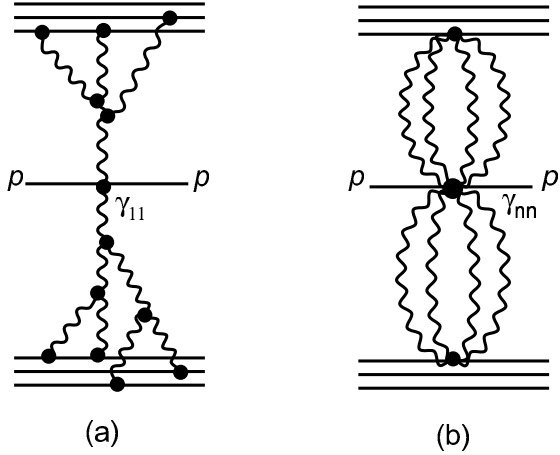}
\caption{Multipomeron diagrams for inclusive production:
(a) - diagrams with fusion and splitting of parton chains;
(b) - diagrams with rescattering of $n$ parton chains.
 }
\label{mult pom}
\end{center}
\end{figure}
Effects of fusion become noticeable already at RHIC energies, resulting
in reduction of secondary density about two times as compared with the eikonal
approximation \cite{kaidalov}. This effect is mainly due to diagrams with
three-pomeron interaction (Fig.\,\ref{mult pom}a). But due to their sign-alternating
character a profile of the amplitude becomes more smooth and the azimuthal asymmetry
becomes even smaller because of smaller gradients of parton density.

Different situation is realized for diagrams with chain rescattering
(Fig. \ref{mult pom}b) which give positive contributions to inclusive cross section.
Even if the total inclusive cross section changes slightly due to such rescatterings,
its anisotropic part may be strongly increased. Indeed, the expression
\eqref{conv0} for inclusive cross section will contain in case of $n$ interacting
chains a convolution $T_A^n\otimes T_B^n$ instead of $T_A\otimes T_B$.
Azimuthal asymmetry is determined by density derivatives (see \eqref{v2T}), therefore
an extra factor $\langle n(b) \rangle^2$ will appear in expression for $v_2$.
The number of interacting pomerons at RHIC energies is not large
($\langle n(b) \rangle \sim 2$ for central collision), but increase of $v_2$
can be considerable because the effect is quadratic in $n$ (see Fig.\ref{num1}
in section \ref{numerical}). Note that $\langle n(b) \rangle$ increases with atomic
number $A$ and changes the $A$ dependence for $v_2$ also.
The same mechanism may give an increase of $\langle p_t^2 \rangle$.

\section{Numerical estimates}
\label{numerical}

Let us present examples of numerical estimates of elliptic flow
for the exponential parametrization \eqref{gamma_exp} of the
inclusive vertex. They are illustrative only. We intentionally do
not fit experimental data because, first, little quantitative
information on the inclusive vertex $\gamma$ exists and, second,
it is necessary to have more reliable model for nuclear
multipomeron effects. In this publication we aim to demonstrate
that qualitative features of elliptic and directed flows peculiar
to the mechanism under discussion are in agreement with
experimental observations. In particular, as it is seen from
Eq.\,\eqref{v2T} the value of  $v_2(p_t,b)$ grows with increase of
both transverse momentum $p_t$ and impact parameter $b$.
\begin{figure}[h]
\begin{center}
\includegraphics[height=80mm]{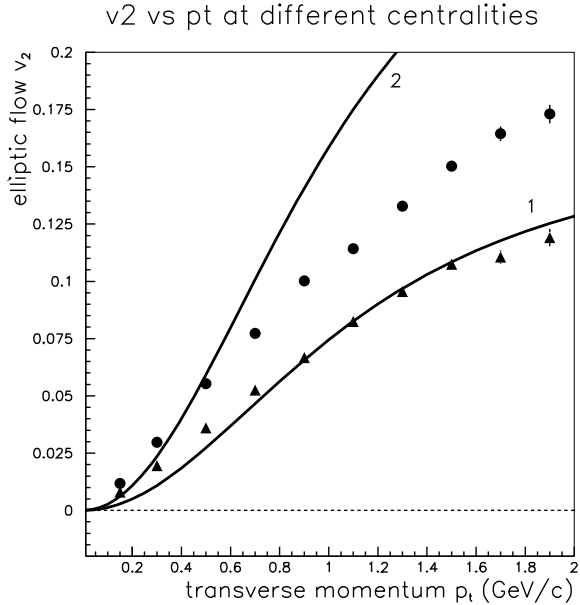}
\caption{Elliptic flow $v_2$ as a function of transverse momentum
$p_t$ in $Au-Au$ collisions for two centrality values:
$C=13\%$ ($b\sim 5.6~\rm{fm}$) -- curve 1;
$C=27.5\%$ ($b\sim 7.9~\rm{fm}$) -- curve 2.
}
\label{num1}
\end{center}
\end{figure}
\begin{figure}[h]
\begin{center}
\includegraphics[height=80mm]{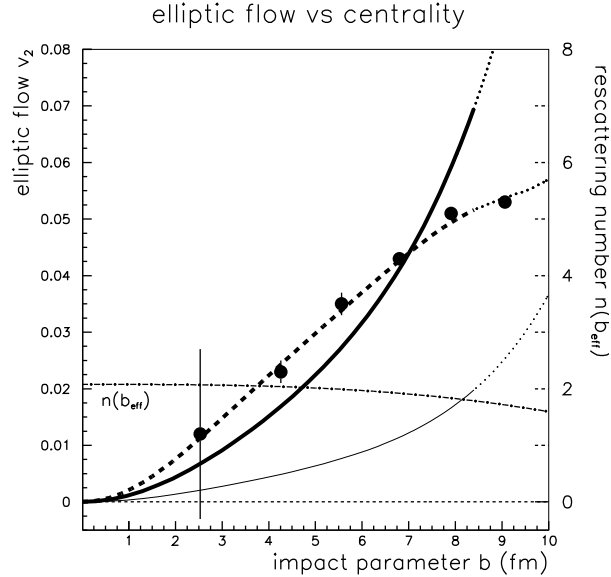}
\caption{ Elliptic flow $v_2$ as a function of impact parameter
$b$ at $p_t=0.5~GeV/c$. Calculations for $b > 8~\rm{fm}$ are shown
with dotted curve. Thick solid line corresponds to calculations at
$\epsilon=0.4$. Thin solid line corresponds to calculations at
$\epsilon=0.1 $, and dotted line -- at $\epsilon=0.1$ multiplied
to the mean rescattering number squared $n(b_{eff})=n(0.65\, b)$.
The dashdotted line at this picture gives  $n(b_{eff})$ as a
function of $b$ (the right scale). } \label{num2}
\end{center}
\end{figure}
\begin{figure}[h]
\begin{center}
\includegraphics[height=80mm]{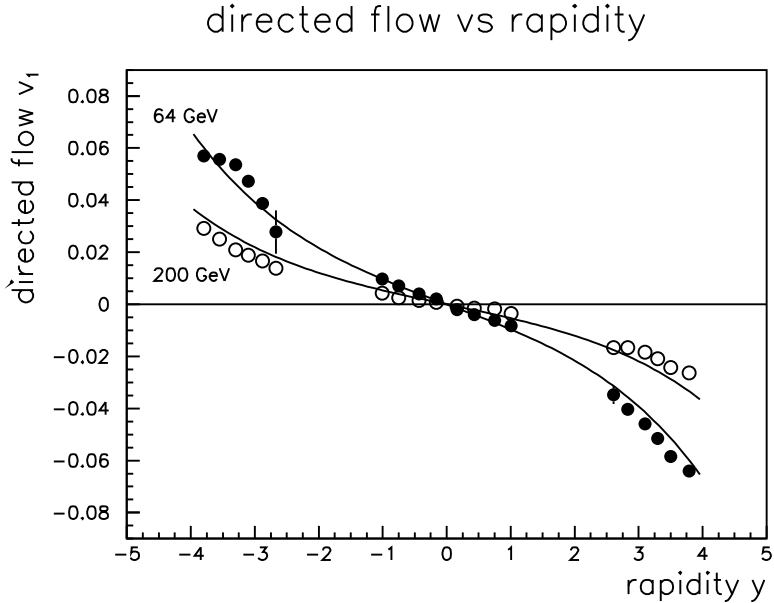}
\caption{Directed flow $v_1$ as a function of rapidity $y$.
}
\label{num3}
\end{center}
\end{figure}

It is shown on Fig.\,\ref{num1} $p_t$-dependence of elliptic flow for two values
of centrality. Calculations were made for two different values of centrality
at $\epsilon = 0.4$ for the standard Woods-Saxon parameterization of nuclear density.
Note that for large $p_t$ the growth of $v_2$ is stopped due to higher order of $\epsilon$
though it is not quite justified to use Eq.\eqref{gamma_exp} in high $p_t$ region.
Experimental data are taken from \cite{STAR0206001}.
The too fast increase of theoretical predictions with centrality may be related
to a neglect of multipomeron diagram contribution (see below).

Fig.\ref{num2} presents the $b$-dependence of elliptic flow. According to Eq.\,\eqref{v2T}
$v_2$ grows as $b^2$ (thick solid line, $\epsilon = 0.4$) in contradiction with
the experimental behavior at very large $b$ when nuclei only touch each other.
Let us note here that in the case of very large impact parameter values
where nuclear densities are quite small an accuracy of the model is low. This is why
calculation results for $b > 8~\rm{fm}$ are shown with dotted curves. The dashed curve
corresponds to the simplest account of nuclear effects due to inter-pomeron interactions
-- calculations at smaller parameter value $\epsilon=0.1$ (thin solid line) are
multiplied to the factor $n(b_{eff})^2$. The function $n(b_{eff})$ gives an effective
number of interacting pomeron fluctuations
\footnote{
~ The function $n(b)$ was estimated in the eikonal approximation with effective
parton-nucleon cross section value $\sigma\sim 10\text{~mb}$. The mean position of
interacting nucleon in a nucleus was taken as $b_{eff}=0.65\, b$.}%
. It is also presented at the figure (dash-dotted line) and one
can see that the maximal number of pomeron interaction is about 2
for central collisions. Experimental data are from
\cite{STAR0206001}.

Dependence of the directed flow $v_1$ on rapidity $y$
calculated at $\delta=0.1$
for two energies (200 and 64 GeV) is presented on Fig.\,\ref{num3}.
It is antisymmetric and concentrated at the edges of the rapidity interval rapidly
decreasing to the middle of it.
The experimental data corresponding to centrality 30 - 60 \%
are taken from ref.\cite{STAR0807}.

\section{Conclusions}
\label{conclusion}

The main aim of this paper is to draw attention to the mechanism of appearance
of azimuthal correlations between produced particles related to specific parton
correlations in initial states of colliding nuclei,
and which is not caused by particle interactions in the final state.

The mechanism under discussion is connected with correlation
between parton transverse momentum and its transverse position inside nucleus.
Transverse momenta of partons in incoming nuclei (before collision)
are already partially aligned along transverse nuclear radiuses.
The non-central collision of nuclei selects the asymmetric overlapping part of the fast
nucleus resulting in  $\p_t-\b$ correlations for produced particles.

We discussed this phenomenon using the simplest reggeon diagram in order to emphasize
the fact of its existence.
For quantitative description of asymmetries it is necessary to account for a more
complicated mechanism related to interactions between different chains.
We emphasized that this mechanism can substantially increase elliptic flow and modify
its atomic number dependence. Role of multipomeron interactions describing these effects
increases with energy and is important at RHIC energies. This mechanism influences
dependence of asymmetries on centrality, transverse momenta and rapidity.

Evidently the final state interaction can also contribute to
azimuthal asymmetry of produced particles. In order to obtain an information on properties
of hadronic matter at high temperatures and densities it is crucial to separate
effects due to initial and final state interactions. In particular it is important
to carry out a detailed investigation of azimuthal correlations of particles
produced in nucleon-nucleon and nucleon-nucleus interactions. Such studies will be
possible in high-multiplicity events at LHC.

\emph{Acknowledgements.}
We are grateful to L.V.~Bravina, Yu.A.~Simonov and E.E.~Zabrodin for stimulating
interest and discussions.
The work was supported by RFBR grants 06-02-17012, 08-02-00677, 06-02-72041-MNTI,
and SSh 843.2006.2.


\begin{thebibliography}*
%
\bibitem{Olli92}
  J.-Y. Ollitrault,
  Phys. Rev. D {\bf 46}, 229 (1992).
%
\bibitem{Olli93}
  J.-Y. Ollitrault,
  Phys. Rev. D {\bf 48}, 1132 (1993).
%
\bibitem{voloshin96}
S.~Voloshin, Y.~Zhang, Z. Phys. C {\bf 70} (1996) 665.
%
\bibitem{Sorge97}
  H. Sorge,
  Phys. Rev. Lett. {\bf 78}, 2309 (1997).
\bibitem{Voloshin02}
S.A. Voloshin,
Nuclear Physics A, {\bf 715},  379 (2003).
%
\bibitem{phenixflow}
S.~S.~Adler {\it et al.}, PHENIX Collaboration, Phys.~Rev.~Lett.
{\bf 91},182301 (2003).
%
\bibitem{phobosflow}
M.~Tonjes (for the PHOBOS Collaboration),
J.~Phys.~G. {\bf 30}, S1225 (2004).
%
\bibitem{starflow}
J.~Adams {\it et al.}, STAR Collaboration, Phys.~Rev.C
{\bf 72}, 014904 (2005).
\bibitem{AGK}
V.A.~Abramovsky, V.N.~Gribov, O.V.~Kancheli,
Yad.~Fiz. \textbf{18}, 595 (1973), Sov.~J.~Nucl.~Phys. \textbf{18}, 308 (1974).
\bibitem{kaidalov}
A.B.~Kaidalov, Phys.~Usp. \textbf{46}, 1121 (2003),
Usp.~Fiz.~Nauk \textbf{46}, 1153 (2003).
\bibitem{STAR0206001} STAR collaboration, Phys.~Rev. C \textbf{66}, 034904 (2002).
\bibitem{STAR0807} STAR collaboration, nucl-ex/0807.1518.
\end{thebibliography}
\end{document}